\documentstyle[prl,aps,twocolumn]{revtex}

\begin{document}
\title{{\bf Surfactant effect in heteroepitaxial growth. The Pb - Co/Cu(111) case}}
\author{Liliana G\'{o}mez}
\address{Facultad de Cs. Exactas e Ingenier\'{i}a -\\
Universidad Nacional de Rosario -\\
Instituto de F\'{i}sica Rosario, 2000-Rosario, Argentina.}
\author{Julio Ferr\'{o}n}
\address{INTEC-FIQ, CONICET - Universidad Nacional del Litoral, \\
3000-Santa F\'e, Argentina.}
\date{\today}
\maketitle

\begin{abstract}
A MonteCarlo simulations study has been performed in order to study the
effect of Pb as surfactant on the initial growth stage of Co/Cu(111). The
main characteristics of Co growing over Cu(111) face, i.e. the decorated
double layer steps, the multiple layer islands and the pools of vacancies,
disappear with the pre-evaporation of a Pb monolayer. Through MC
simulations, a full picture of these complex processes is obtained. Co
quickly diffuses through the Pb monolayer exchanging place with Cu atoms at
the substrate. The exchange process diffusion inhibits the formation of pure
Co islands, reducing the surface stress and then the formation of multilayer
islands and the pools of vacancies. On the other hand, the random exchange
also suppress the nucleation preferential sites generated by Co atoms at Cu
steps, responsible of the step decoration.
\end{abstract}

\vspace{1cm} PACS numbers: 68.35.Fx, 68.55.-a, 81.10.-h


\narrowtext

\section{Introduction}

The startling, and sometimes surprising physical properties observed in the
new series of artificial materials, has catapulted an intensive work around
the world. The experimental conditions needed to obtain atomically clean and
nearly perfect flat substrates are achieved in most surface labs. Currently,
almost any atom can be deposited over almost any surface, allowing the
creation of phases that are not found in nature. But, the way the atoms
diffuse over the terrace (intra layer diffusion) and the ability to overcome
terrace steps (inter layer transport) before binding other walking atoms,
will determine the morphology of the growing film. Being intermixing and
surface roughness development the most common artifacts in the preparation
of these materials, the obtaition of flat films through layer by layer
growth (LbL) is mostly the exception than the rule. External parameters,
like substrate temperature and evaporation rate among the most usual, and
internal ones, like interaction potentials, Ehrlich-Schwoebel barrier,
diffusion type, surface orientation and so on, are crucial in determining
the type of film growth \cite{besem,vander,s2,s3}. On the other hand, the
evaporation or adsorption of some materials, known as surfactants, previous
to the metal deposition has proved to promote LbL growth \cite
{O-sur,camarero1,camarero2}. A perfect example of this behavior is the
growth of Cu over Cu(111). In fact, while in absence of surfactant Cu grows
in a 3D fashion over Cu(111), after the pre evaporation of a monolayer of
Pb, LbL growth in a wide range of temperatures and evaporation rates can be
obtained. \cite{prl_exchange} In a recent work, we showed that the monolayer
of Pb changes the diffusion mechanism of Cu over the surface from hopping to
atomic concerted exchange, turning the diffusion over Cu(111) surface like
over the Cu(100) one, where LbL growth is usually obtained \cite
{prl_exchange}.

The growth of Co over Cu(111), on the other hand, is a good example of
hetero-epitaxial non LbL growth. Multiple height island formation, step
decoration, vacancy (pools) islands and Cu-Co intermixing are common
features in this system \cite{experimental,prlco}. In a recent work, we
showed that this behavior depends on the adsorbate - substrate interaction,
and it should be awaited any time materials of these types are put into
contact \cite{surf}. As in the case of Cu/Cu(111), the pre-evaporation of a
monolayer of Pb acts as a surfactant promoting 2D growth, and allowing the
fabrication of good quality magnetic multilayer. In this work we present
results, based on MonteCarlo simulation about the effect of Pb as surfactant
for heteroepitaxial growth, in particular we analyze the case Co/Cu(111)

\section{Computational details}

The interaction potentials employed in our MC simulations have been obtained
using the second-moment approximation of the tight-binding scheme (TB-SMA) 
\cite{tomanek}. They include a short-range, repulsive pair potential plus a
long-range, many-body contribution based on a tight-binding description of
the electronic structure \cite{ducastelle}. They have been successfully used
previously to study the properties of transition metal surfaces \cite
{gupta,mottet}, noble metals \cite{guillope} and microclusters \cite
{tomanek,diep,sawada}. This potential will be used for the six involved
interactions, i.e. Co-Co, Pb-Pb, Cu-Cu, Co-Cu, Co-Pb and Cu-Pb with their
corresponding physical parameters.\cite{surf} The potential at the $i-th$
atom is given by

\begin{eqnarray}
V_i(r_{ij})& = &U \{ A \sum_{j} \exp [-p(r_{ij}-r_{0})]  \nonumber \\
& &\hspace{0.5cm}- [\sum_{j} \exp[-2q(r_{ij}-r_{0})]]^{1/2}\}
\end{eqnarray}

where $r_{ij}$ is the distance between the two atoms at $r_{i}$ and $r_{j}$,
respectively, $r_{0}$ is the nn distance of the bulk crystal, $U$, $A$, $p$
and $q$ are parameters. In this paper, the sums are performed over all atoms
within a sphere of radius equal to 3.2 times the Cu nn distance. We have
checked that including interaction between further atoms does not alter the
conclusion of this paper. The many-body interactions are included in the
potential through the square root of the second sum. This term takes into
account the essential band character of the metallic bonds \cite
{ducastelle,sutton,chadi}. The value of $A$ is determined minimizing the
cohesive energy of the bulk crystal with nn distance $r_{0}$, while the
values of $U$, $p$ and $q$ are determined in such a way that the bulk
cohesive energy and the bulk modulus calculated using (1) and the
experimental value of $r_{0}$ are in good agreement with the experimental
bulk cohesive energy and bulk modulus. The values $p=9/r_{0}$ and $q=3/r_{0}$
are found to be appropriate for the surface \cite{gupta}. The interactions
between the different types of atoms are determined by sets of six
parameters $\{U_{i},p_{i},q_{i}\}$; $i=$ 1 stands for Cu-Pb pairs, $i=$ 2 for
Pb-Pb, and so on. {$r_{0_{i}}$} are the bulk nearest-neighbor (nn)
distances. The validity of this approximation has been tested in previous
studies with this combination of elements. The ordered structures and
melting behavior of Pb overlayers on Cu(100) have been successfully
reproduced \cite{gomez}. In recent studies, the bulk properties of Pb,
including the phonon spectrum, have been calculated giving excellent
agreement with the experimental results \cite{cleri,dobry}. The Cu(111)
sample used in our calculations consisted of 6 layers with 128 Cu atoms in
each; to study the surfactant effect, a compact layer of 72 Pb atoms was
added to the upper surface of the slab. The two lowest Cu layers were frozen
to simulate the bulk, while all the remaining atoms in the sample were
allowed to move. Most of the calculations have been performed at a sample
temperature of 540 K in order to speed up the diffusion processes and
improve the statistics.

\section{Results and discussion}

As different as the growth processes of Cu and Co over Cu(111) is, the
effect of Pb as surfactant recognizes the same root. Cu grows over Cu(111)
in 3D from the beginning. This is so due to the low intralayer diffusion
barrier compared to the interlayer one (Ehrlich-Schwoebel barrier).
 The presence of a
monolayer of Pb changes the diffusion mechanism from hopping to exchange,
making both barriers of the same order. In absence of any other effect, this
is enough to promote 2D growth \cite{prl_exchange}. The heteroepitaxial
growth of Co over Cu(111) is by far more complex. The growth of Co over nude
Cu presents the formation of double atomic height decorated steps,
multilayer islands and pools of vacancies \cite{prlco}. Through ion
scattering spectroscopy \cite{24} and chemical titration \cite{25} it has been shown
that these islands are actually formed by a mixture of Co and Cu atoms. 
 In a recent paper \cite{prlco} we showed that the diffusion
mechanisms, the larger surface energy of Co as compared to Cu, and the Co-Co
and Co-Cu interaction potentials are the responsible of these features. Over
the Cu terrace Co diffuses through hopping. In this way, the adatom
nucleation gives place to the formation of pure Co island. Once the Co
islands reach a critical size, they explode forming multilayer alloyed
islands, and leaving pools of vacancies. On the other hand, the interlayer
mass transport is produced, as in the case of Cu-Cu, through the exchange
mechanism. But, in this case the process gives place to alloyed steps that
act as nucleation sites for the diffusing Co adatoms, forming the double
atomic height decorated steps.

The question to answer is then if the effect of Pb on Co/Cu(111) growth is
similar to the homoepitaxial case. And if the change in the diffusion
mechanism is enough to suppress all the different features appearing in
Co/Cu(111), promoting 2D growth. This is just the case; in Fig.1 we show the
evolution of a Cu (left panel) and a Co adatom (right one) over a Cu(111)
surface pre - covered with a Pb monolayer, obtained through MC simulations.
In the upper panel the evolution of the z coordinate (normal to the surface)
is depicted vs. MC steps. As we know, a drop in the Z coordinate is a
fingerprint of a concerted exchange process. Thus, the first drop observed
in both cases represents the exchange of Cu (Co) atom with the surfactant.
This exchange is very fast and it is the responsible of the floating of the
Pb monolayer. The second drop in z coordinate (not always occurring within
the elapsed time analyzed) corresponds to the CoCu (CuCu) exchange. As we
showed for Cu \cite{prl_exchange}, the Co diffusion over Pb/Cu(111) system
occurs below the Pb layer and through the exchange mechanism. We should note
that although Co and Cu adatom diffusion below Pb are quite similar, they
are not identical as well. In the lower panel of Fig.1 we show a couple of
snapshots of the MC diffusion simulations. We can see that substrate atoms
involved in the concerted exchange mechanism are no the same in both cases.
While CuCu exchange involves next nearest neighbors, the process in the CoCu
case is between nearest neighbors. This difference is maintained for the
following steps, always in the vicinity of the implanted Co atoms. Thus,
although the adatom diffusion below the surfactant layer is still through
concerted exchange, the presence of Co as impurity modifies the nature of
the exchanging mechanism. This difference with Cu/Cu exchange may have
important physical consequences. In fact, we have already shown\cite{prlco}
that the interaction CoCu produces a dip binding well, locating the evolving
atom in the near defect zone. Thus, the hopping mechanism will be more
restricted than in the case of Cu/Cu. In fact, we are resembling the
situation responsible of the step decoration \cite{prlco}, but in the middle
of the terrace. Additionally, the probability of a Cu Co reverse exchange,
i.e. extracting the Co atom again, will be larger than in a ''long jump
exchange ''like in CuCu case, increasing in this way the probability of
reacher Co island formation.

The change in the diffusion mechanism, from hopping to exchange is, as in
the case of Cu over Cu(111), enough to explain all the features of the
surfactant effect over the heteroepitaxial growth. In fact, through the
exchange mechanism, very quickly the diffusing atoms turn to be Cu ones, and
no pure Co island could be formed. The alloyed islands lowers the surface
energy affecting the critical size instabilities we found for Co over nude 
Cu
\cite{prlco}. The non-adiabatic production of these alloyed islands, coming
from the unstable pure Co islands, was the responsible of the appearance of
the pools of vacancies and the multiple height island, which are hardly
formed in the new scenario. The supression of the decorated steps is based on
the same mechanism. Since the diffusing atoms are mostly Cu ones, the CoCu
alloyed steps, and thus the preferential nucleation sites located there, are
not formed. The interlayer diffusion is still through concerted exchange,
but the morphology of the steps does not change, since the exchanging
process is mostly CuCu.

We have established that the 1 surface energy is the reason for the
explosion of the Co islands., i.e. pure Co islands over Cu(111) surface are
naturally unstable.  The exchange diffusion mechanism will tend, as we
discussed in the previous paragraph, to the formation of mixed Co-Cu
islands, lowering the surface energy. But, we have not shown still to what
extent this effect is sufficient to suppress the double height island
formation. In fact, it could occur that the double height islands are still
formed under different experimental conditions. For instance they may form
with larger critical size, or they could depend on either coverage or
evaporation rate. Actually, we have found that the CoCu islands are stable
below the surfactant layer in all cases, even for the pure Co islands, due
to the stabilizing effect of the Pb layer. In Fig. 2 we show this effect
through a couple of snapshots of the evolution of a Co island contains 28 
atoms with and
without the Pb covering layer (Pb atoms are not shown for clearness). The
first (two upper) panel show the initial situation, the same for the with
and without surfactant case. The size of the studied island is quite larger
the critical one \cite{prlco}. In the second panel, corresponding to Co over
Cu(111) ( middle) the complete process of multilayer mixed island and the
appearance of vacancy pools is apparent. In the lateral sight (the right
one) the mixed composition of the island is evident. In the upper vision
(left one), the mobility of the island is also apparent. In the third lower
panel,\thinspace the surfactant assisted case, the Pb stabilized island is
shown. Thus, the Pb modification of the diffusion mechanism not only prevent
the formation of pure Co islands, promoting 2D growth, but also, in the
remote, but statistically possible case that either Co pure or Co rich
island are formed, the Pb layer acts stabilizing them.

\section{Conclusions}

Through MC simulation, we have studied the growth of Co over Cu(111) in the
presence of a surfactant (Pb) layer. We found that the surfactant changes
the diffusion mechanism of Co over the surface from hopping to concerted
exchange. An important difference with the previously reported surfactant
effect on Cu/Cu growth, is that the exchange process  involves nearest
neighbor atoms. This effect reduces even more the hopping mechanism, and
increase the probability of reverse exchange, only important in hetero
epitaxy. The surface diffusion through exchange diminishes the probability
of  pure Co islands formation, which are the responsible of the multiple
height and vacancy island formation. In the same way, the exchange mechanism
suppresses the formation of the binding alloyed steps, responsible of the
step decoration. Additionally, the coverage of the Co by Pb atoms diminishes
the effect of the different surface energy between Cu and Co stabilizing
even the pure Co island. This effect turns irrelevant the non zero
probability of the formation of pure Co island.

\begin{figure}[tbp]
\vskip .5 truecm
\vskip .5 truecm
\caption{Surface diffusion of Cu and Co over Cu(111) pre-covered with a
monolayer of Pb obtained by means of MC. First column, x, y and z
evolution of Cu over
Pb/Cu(111). The first drop corresponds to the initial Cu Pb exchange;
Second column: as
in a, for Co adatom diffusion. 
Lower panel: top views showing the exchange
between the Cu substrate atoms (gray clear - gray obscuro circle) and the subsurface 
evolving adatom,
 i.e. Cu
or Co (black circle). Note that while the exchange Cu/Cu is with the 
next-nearest neighbor,
in the Co case is with the nearest one. The Pb atoms are not shown in the last
frames for clarity}
\label{fig1}
\end{figure}


\begin{figure}[tbp]

\vskip .5 truecm
\vskip .5 truecm
\caption{Co comparative island stability over Cu(111) with and without a
surfactant layer. Upper (first column) and lateral (second column) view.
Upper panel: Original situation for a Co island over the critical size. 
Middle panel: snapshot of the Co island over nude Cu, the multilayer island
formation, together with the pools
of vacancies can be observed.
Lower panel: the evolution of the same Co
island, under the surfactant layers is shown. The island, as well as the
substrate, shows some kind of corrugation, but for comparative times, the
island stability is proved}
\label{fig2}
\end{figure}

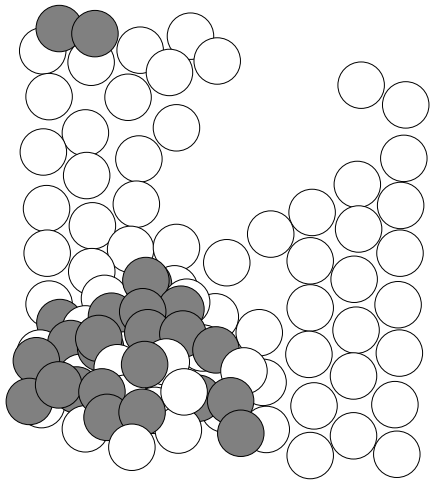
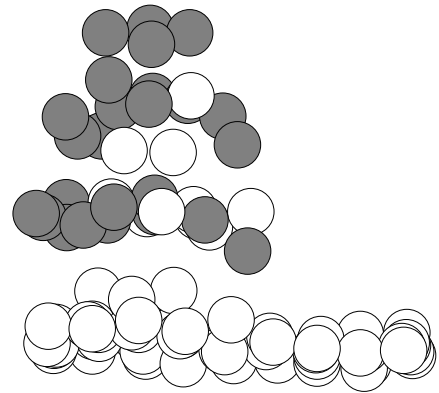
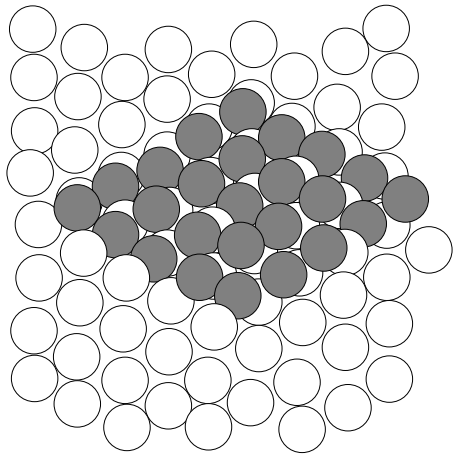
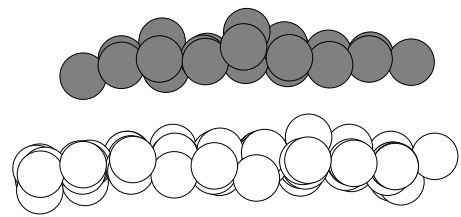
\end{document}